\documentclass[aps,prl,twocolumn,superscriptaddress]{revtex4}

\usepackage{times}
\usepackage{color}
\usepackage{amsmath, amsthm, amssymb,  mathrsfs, epsfig}
\begin{document}

\title{A Blueprint for a Topologically Fault-tolerant Quantum Computer}

\author{Parsa Bonderson}
\affiliation{Microsoft Research, Station Q, Elings Hall, University of California, Santa Barbara, CA 93106}
\author{Sankar Das~Sarma}
\affiliation{Microsoft Research, Station Q, Elings Hall, University of California, Santa Barbara, CA 93106}
\affiliation{Department of Physics, University of Maryland, College Park, MD 20742}
\author{Michael Freedman}
\affiliation{Microsoft Research, Station Q, Elings Hall, University of California, Santa Barbara, CA 93106}
\author{Chetan Nayak}
\affiliation{Microsoft Research, Station Q, Elings Hall, University of California, Santa Barbara, CA 93106}
\affiliation{Department of Physics, University of California, Santa Barbara, CA 93106}

\maketitle

{\bf The advancement of information processing into the realm of quantum mechanics promises a transcendence in computational power that will enable problems to be solved which are completely beyond the known abilities of any ``classical'' computer, including any potential non-quantum technologies the future may bring. However, the fragility of quantum states
poses a challenging obstacle for realization of a fault-tolerant
quantum computer. The topological approach to quantum computation proposes to surmount this obstacle by using special physical
systems -- non-Abelian topologically ordered phases of matter --
that would provide intrinsic fault-tolerance at the hardware level.
The so-called ``Ising-type'' non-Abelian topological order is likely to be physically realized in a number of systems, but it can only provide a universal gate set (a requisite for quantum computation) if one has the ability to perform certain dynamical topology-changing operations on the system.
Until now, practical methods of implementing these operations were unknown.
Here we show how the necessary operations can be physically implemented for Ising-type systems realized in the recently proposed superconductor-semiconductor and superconductor-topological insulator
heterostructures. Furthermore,
we specify routines employing these methods to generate
a computationally universal gate set. We are consequently able to provide a schematic blueprint
for a fully topologically-protected Ising based quantum computer using currently
available materials and techniques. This may serve as a starting point
for attempts to construct a fault-tolerant quantum computer,
which will have applications to cryptanalysis, drug design,
efficient simulation of quantum many-body systems,
solution of large systems of linear equations,
searching large databases, engineering future quantum
computers, and -- most importantly -- those applications
which no one in our classical era has the prescience to
foresee.}

The vulnerability of quantum computers to errors
can, theoretically, be overcome by quantum error
correction~\cite{Shor95}.
However, fault-tolerance is difficult
to engineer in practice because quantum
error correction protocols introduce more errors than
they correct unless the error rate
is very small to begin with~\cite{Aharonov97,Knill98}.
The correctable error threshold is estimated to be
$\stackrel{<}{\scriptstyle \sim} 10^{-3}$
per computational gate operation~\cite{Aliferis07}, which is quite difficult to achieve.
An alternative approach is to implement
fault-tolerance at the hardware level.
This approach can be realized, in principle,
by a non-Abelian topological phase of matter~\cite{Kitaev97,Freedman98,Preskill98,Freedman02a,Nayak08}.

\begin{figure}[b!]
\centering
\includegraphics[width=3.5in]{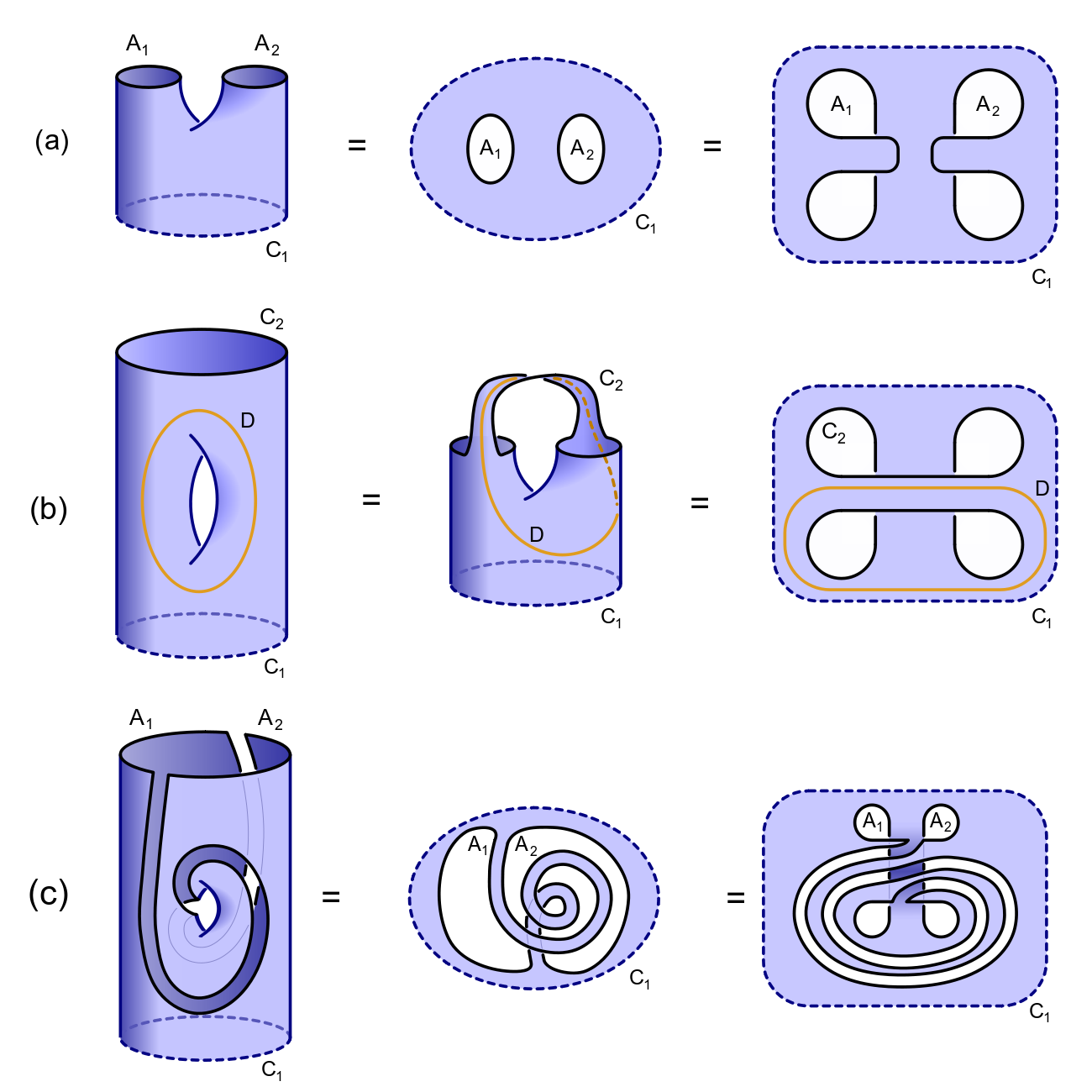}
\caption{Configurations for universal topological quantum computing with Ising-type systems. Each row contains three geometrically different realizations of topologically equivalent surfaces.}
\label{fig:topology}
\end{figure}

Topology is a branch of mathematics which
focuses on those properties of
mathematical spaces, e.g. surfaces, which are invariant under continuous
deformations. For instance, the three surfaces displayed in Fig.~\ref{fig:topology}(a) are topologically equivalent, since each can be obtained from any other by stretching it like a sheet of rubber without cutting, puncturing, or gluing. Similarly, the three surfaces in Fig.~\ref{fig:topology}(b) are topologically equivalent, as are the three in Fig.~\ref{fig:topology}(c). In fact,
the surfaces in Fig.~\ref{fig:topology}(a) and \ref{fig:topology}(c) are also topologically equivalent,
but with very different geometries. It is worth the
reader's time to visualize the equivalencies.

Topology is an essential aspect of many physical systems, e.g. vortices in superfluids and superconductors, defects in liquid crystals, magnetic domain structures, and van Hove singularities in crystal spectra.
One of the most remarkable developments in physics
in the last 30 years is the discovery of \emph{topologically ordered}
phases of matter~\cite{Wen90a}, which are extreme examples of
topology governing physics: {\it all} low-energy,
long-distance properties of the system are unaffected by
any local perturbations. These phases
were initially discovered in the quantum Hall regime~\cite{Tsui82,Laughlin83,Halperin84}.
As the theoretical description of such phases --
topological quantum field theories (TQFTs)~\cite{Witten89} -- was
developed, it was later realized that ordinary superconductors
are simple topological phases~\cite{Hansson04} and that some more
exotic superconductors, such as chiral $p$-wave
superconductors, may be more intricate
topological phases~\cite{Read00}.
Although the subject is relatively new, many models of topological phases
have been analyzed theoretically,
motivated largely by the possibility of topological quantum computation (TQC), see e.g.~\cite{Nayak08} and references therein.

\begin{figure}[t!]
\centering
\includegraphics[width=3.5in]{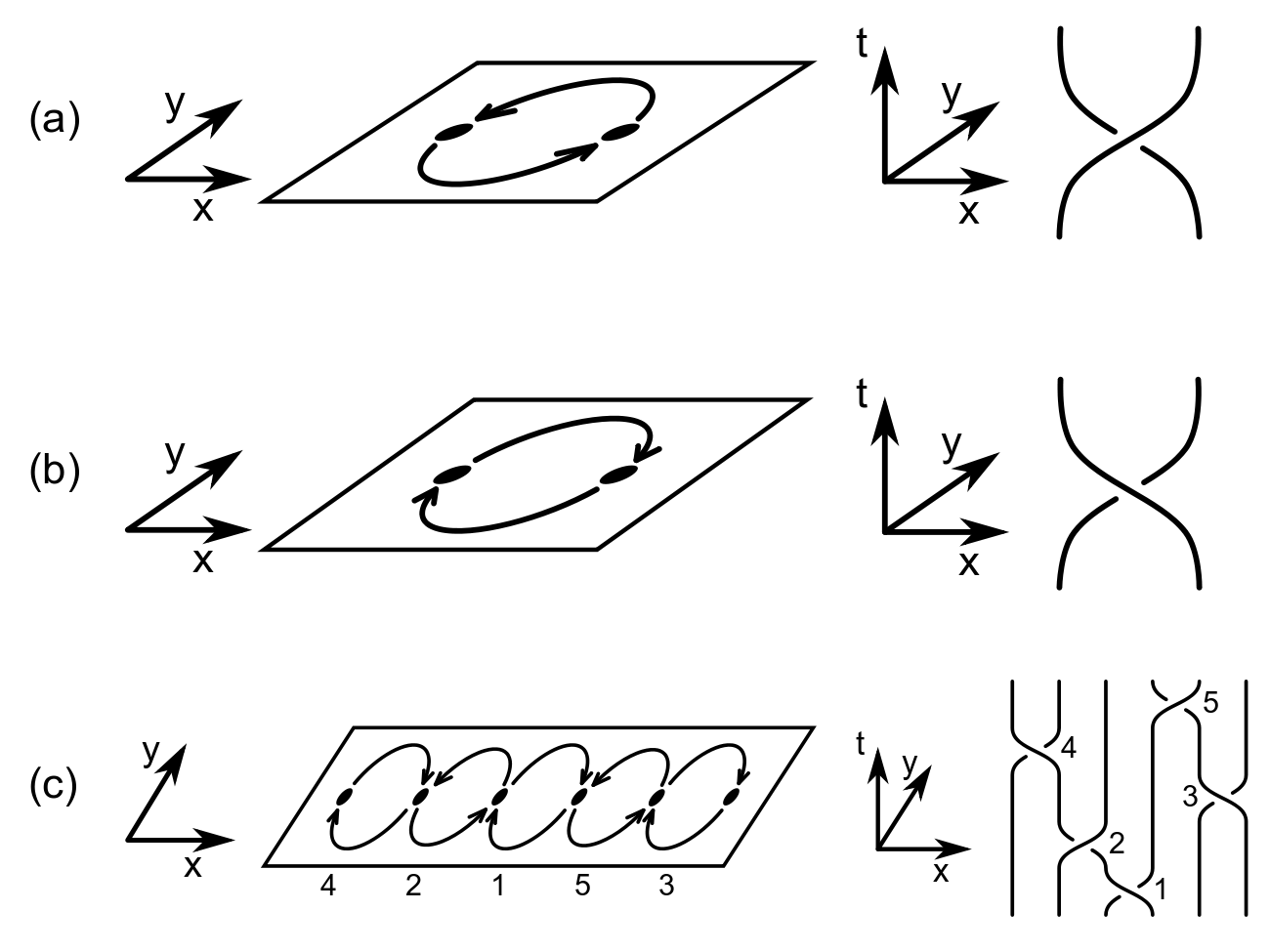}
\caption{Braiding anyonic quasiparticles. The spatial and spacetime trajectories
of (a) counterclockwise exchange, (b) clockwise exchange, and (c) a series of exchanges (occurring in the numbered order) resulting in a more complicated braid of worldlines.}
\label{fig:braiding}
\end{figure}

Topological phases are characterized by the exotic properties of their quasiparticle excitations, which are called anyons.
For many applications, anyons
are the correct way to study the system.
Electrons, from this perspective, are higher-energy
composites, and looking at the system in terms of their
behavior would be like doing electrical engineering with quarks.
The region of the system which supports the topological
phase and, therefore, anyons
is referred to as a {\em topological fluid}.
Since such systems are effectively two-dimensional, the simple interchange of the positions of two quasiparticles has a definite handedness, as shown in Fig.~\ref{fig:braiding}. Consequently, quasiparticles' worldlines (space-time trajectories) form braids in $2+1$ dimensional space-time, and the term ``braiding'' is often used to describe quasiparticle exchange. Different series of interchanges are distinguishable if the resulting worldlines form topologically distinct braiding configurations~\cite{Leinaas77}.
Non-Abelian topological phases are those in which
quasiparticles give rise to a degenerate non-local state space. Braiding operations have the effect of applying unitary transformations (matrices) to this state space~\cite{Goldin85,Fredenhagen89,Froehlich90}. The utility of
such non-Abelian phases of matter for fault-tolerant quantum computation~\cite{Kitaev97,Freedman98,Preskill98,Freedman02a,Nayak08} stems from the fact that quantum information
can be encoded and processed in this non-local state space in a manner that is inherently protected from errors, since it is impervious to the effects of the environment when the temperature is low.
In this regard, braiding is the primary means of generating
computational gates, though it has been shown
that measurements can be used to generate all braiding
transformations~\cite{Bonderson08a}.

Two of the simplest
non-Abelian topological phases are described theoretically
by the so-called ``Ising'' and ``Fibonacci'' TQFTs. The Ising TQFT is
characterized by a type of non-Abelian quasiparticles, called
$\sigma$ quasiparticles, pairs of which share a quantum two-level
system. Thus, each pair of $\sigma$s can be viewed as
a qubit. Superconducting vortices in the structures described in this paper
are predicted theoretically to be $\sigma$ quasiparticles.
There is some experimental evidence~\cite{Dolev08,Radu08,Willett09}
and strong theoretical arguments that charge $e/4$ quasiparticles
(the electrical charge of a $\sigma$ may vary from one
system to another) in the $\nu = 5/2$ quantum Hall
state are $\sigma$ quasiparticles~\cite{Moore91}.
Braiding operations implement $90$ degree rotations~\cite{Nayak96c} in the Hilbert space of these qubits,
which is convenient
for many of the operations needed in typical quantum algorithms, since they are Clifford operators.
However, these unitary transformations form a finite set and only allow one to perform computations
that can also be carried out using a classical computer.

In contrast, the Fibonacci TQFT is characterized by quasiparticles whose braiding operations
can approximate any transformation on the state space to any desired level of precision, and can therefore provide a proper
{\it universal} quantum computer. However,
many simple transformations, such as a NOT gate,
although they can be approximated to
any desired accuracy, require extremely complicated
braids~\cite{Bonesteel05}
(with several orders of magnitude more braiding operations than for Ising). There is not
strong enough evidence for the existence in nature
of Fibonacci anyons for their use
to be anything but a theoretical dream at present.

There is, however, a fortunate loophole in this seemingly
bleak picture: a system in the universality
class of the Ising TQFT can actually be a universal
quantum computer if the gate set obtained from
braiding is supplemented by a $\pi/8$ phase gate
and a Controlled-Z, ${\rm C}(Z)$, gate~\cite{Boykin99}. This can be done in either of two ways: (1) implementing these gates in a topologically unprotected manner~\cite{Bonderson09},
which consequently requires the use of error-correction protocols (though in this case, one can use ``magic state distillation''~\cite{Bravyi05,Bravyi06}, which has a remarkably high error threshold of approximately $0.14$),
or (2) by performing operations that change the topology of the system, which allows these gates to be obtained in a topologically protected manner~\cite{Bravyi00-unpublished,Freedman06a}.

For the implementation of the second route, recent
developments~\cite{Fu08,Sau09,Alicea09} involving what we refer to generically as
``Ising sandwich heterostructures'' (ISHs) are extremely exciting and encouraging. The potential advantages of these ISH systems are:
(1) the temperature scale may be relatively high
(i.e. above 1K), (2) a large perpendicular magnetic field is
not needed, and (3) extremely high-purity specialized materials
are not needed. However, there is a fourth advantage which is the
focus of this paper: constructing systems with non-trivial topology
is quite plausible using sandwich structures, unlike in
quantum Hall devices, as discussed below.

\begin{figure}[t!]
\centering
\includegraphics[width=3.5in]{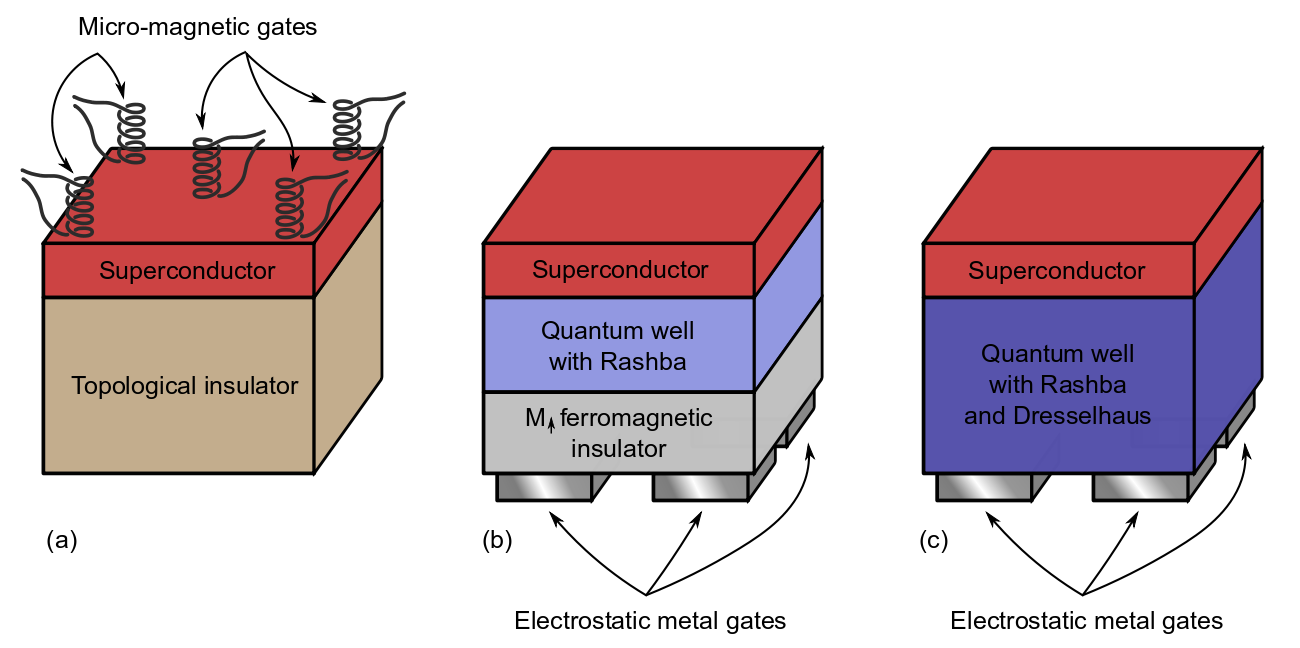}
\caption{Three possible Ising sandwich heterostructures (ISHs):
(a) a superconductor-topological insulator heterostructure~\cite{Fu08},
(b) a superconductor-semiconductor-magnetic insulator heterostructure~\cite{Sau09},
and (c) a superconductor-semiconductor heterostructure~\cite{Alicea09}.
Micro-magnetic gates are added to the top of the
ISH structure in (a) and electrostatic gates to the bottoms
of the structures in (b) and (c). Such individually controllable gates can be used to locally eliminate the topological fluid and manipulate
vortices within the structure immediately beneath/above them.}
\label{fig:ISHs}
\end{figure}

The basic idea behind ISHs, as depicted in Fig.~\ref{fig:ISHs}, is to create an
interface between an ordinary $s$-wave superconductor
and a two-dimensional metal which has the special property
that there is a single fermionic excitation at each wavevector.
This is unlike the situation in normal metals, in which there are
two fermionic excitations at each wavevector (one of each spin).
This special property can be achieved through strong spin-orbit coupling and is guaranteed
for the surface states of a topological insulator~\cite{Fu06}.
Hence, for the ISH in Fig.~\ref{fig:ISHs}(a), one simply uses a superconductor and a topological insulator~\cite{Fu08}.
To generate these conditions using a standard semiconductor, the ISH in Fig.~\ref{fig:ISHs}(b) uses a quantum well with Rashba spin-orbit coupling and a second interface with a ferromagnetic insulator~\cite{Sau09}. Similarly, the ISH in Fig.~\ref{fig:ISHs}(c) generates the desired conditions by using a semiconductor quantum well with Rashba and Dresselhaus spin-orbit coupling and an in-plane magnetic field~\cite{Alicea09}.
Given that there is a surface metal with only a single branch of fermionic excitations, the superconductivity induced in the ISH's surface metal by the adjacent superconductor via the proximity effect is necessarily mathematically equivalent to a chiral $p$-wave superconductor.
Chiral $p$-wave superconductors realize the Ising TQFT
because their vortices support Majorana fermion
bound states \cite{Read00}, which are Ising
$\sigma$ quasiparticles. The vortices can be manipulated
by depleting the topological fluid
with magnetic gates (Fig.~\ref{fig:ISHs}(a))
and/or electric gates (Fig.~\ref{fig:ISHs}(b),(c)).
(The use of the term ``gates'' in reference to these physical electric/magnetic devices
should not be confused with computational gates.)

To date, the best-studied topological insulators
are Bi$_{1-x}$Sb$_x$, Bi$_2$Te$_3$, and
Bi$_2$Se$_3$~\cite{Hasan10}. In the doped semiconductor
approach, InAs, InGaAs, InSb, or even GaAs are possibilities.
The insulating/semiconducting
material in an ISH is not required to be exceptional, e.g. ultra-high mobility, as is required for fractional quantum Hall systems.
However, it is important for it to have a good interface with a
superconductor so that superconductivity can be induced
by the proximity effect. The superconductor does not need to
have any special properties either, but in order to maximize the
proximity effect it should have as large a gap
as possible and to form a clean interface with the
insulating/semiconducting material. Nb is a natural choice
in this regard because it has a large gap and
superconducting proximity effect at Nb-semiconductor
interfaces has already been demonstrated~\cite{Chrestin97,Doh05}.
The large superconducting gap provides the topological
protection for the ISH. This contrasts with the fractional
quantum Hall effect where the gap
$\stackrel{<}{\scriptstyle \sim} 500$~mK
arises from electron-electron interaction and is
easily degraded by imperfections such as non-planarity
and disorder.

\begin{figure}[t!]
\centering
\includegraphics[width=3.5in]{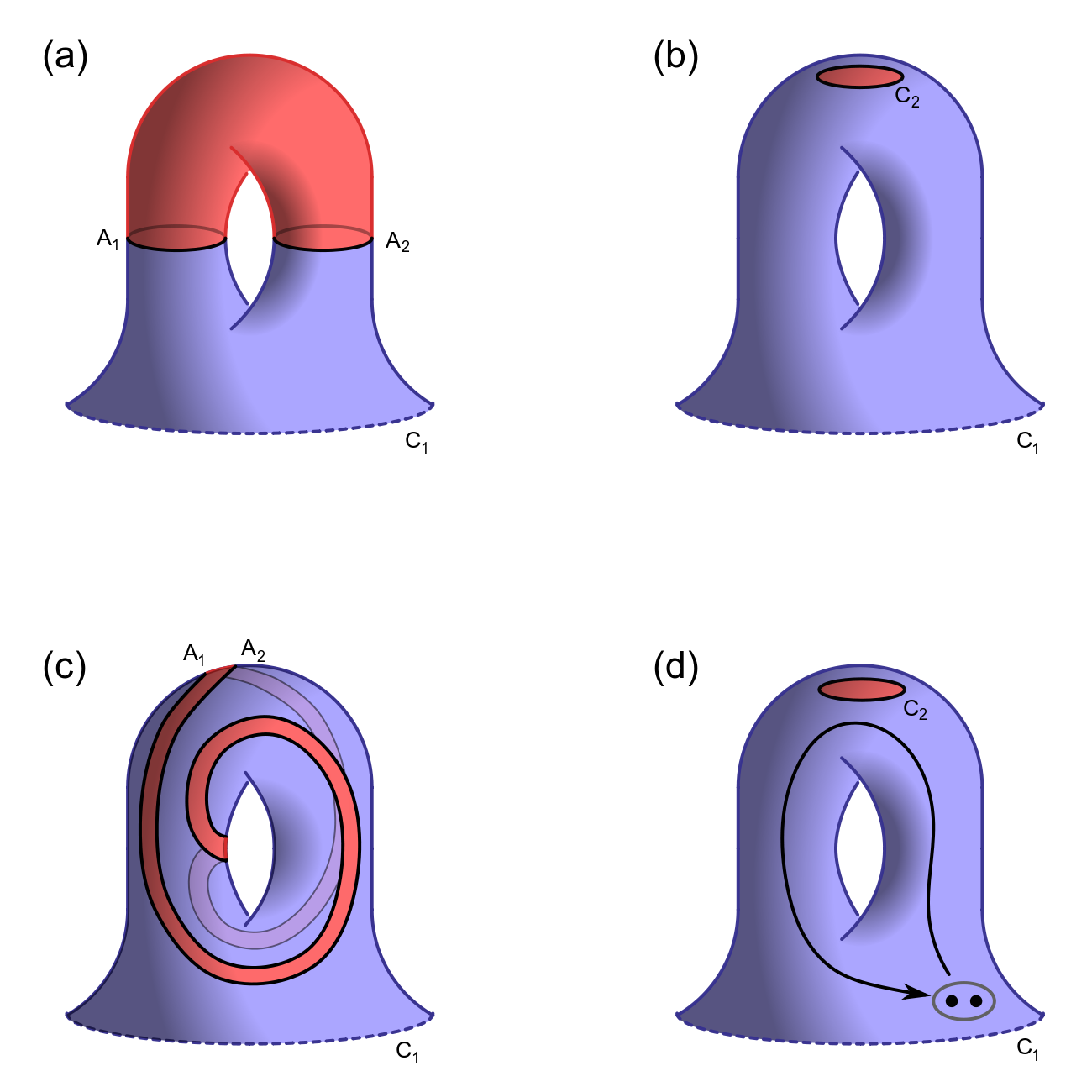}
\caption{Dynamical topology changing (DTC) structures  may be implemented in ISH systems by constructing fixed handle structures (non-trivial topology) in the system with an array of individually controllable depletion gates patterned on the surface of these structures. The topological fluid (blue region) is eliminated
from the ISH below the gates that are turned on (red regions). By controlling an appropriately designed pattern of gates, the topological fluid can be modified to realize the topologies displayed in Fig.~\ref{fig:topology}.}
\label{fig:overpasses}
\end{figure}

While the practical details of TQC implementation in ISHs
will depend on the particular system being used, one can generally categorize the physical operations into four fundamental primitives.
(1) Creation of $\sigma$ quasiparticles.
Flux $hc/2e$ vortices are $\sigma$s
and they are created by threading flux through a depleted
region or hole in ISH.
(2) Measurement of topological charge in a (quasi-)localized region.
We are used to thinking of charge concentrated at a point so it is perfectly natural to think of measuring charge by integrating some emanating flux around a loop. Similarly, topological
charge of a quasiparticle can be detected at a distance by an
interferometric measurement.
This has been studied for the $\nu=5/2$ fractional
quantum Hall effect in theory (see e.g.~\cite{Bishara09} and references therein) and experiment~\cite{Willett09}.
Based on this foundation, interferometry in
ISH systems is rapidly coming under good
theoretical control~\cite{Fu09,Akhmerov09}.
More generally if several quasiparticles are surrounded by a simple closed curve in the topological fluid, a similar measurement will project into a collective charge sector. Surprisingly,
there is a further generalization: any simple closed curve in the bulk --
even if it is caught up in topology and does not simply bound a collection of quasiparticles (for example, the meridian on a hypothetic torus of topological fluid) -- carries a well defined superposition of topological charges which can be probed by the same interferometric set up used in the more conventional case.
(3) Adiabatic transport of anyons, including braiding.
This involves transporting vortices by applying
suitable electric and magnetic fields~\cite{Fu08}.
(4) Deformation of the effective boundary of the topological fluid, including alteration of the fluid's topology (e.g. make or break an overpass), again through electric and/or magnetic gates.
In particular, the ability to change a surface's genus, i.e. the number
of handles (see Fig.~\ref{fig:overpasses} for a genus 1 surface),
is the key idea for making ISHs computationally universal.

In the $\nu=5/2$ fractional quantum Hall effect,
these primitives have been explored by theorists, and even
touched upon in experiments~\cite{Radu08,Dolev08,Willett09},
but non-trivial genus is technologically impossible, putting universal TQC out of reach.
The situation is dramatically different with ISHs, where
we show that genus change is possible.
One only needs to realize two distinct topologies in order to
make Ising anyons universal (though one of these will also
require two different geometries, i.e. Fig.~\ref{fig:topology}(a) and \ref{fig:topology}(c)). These two topologies, shown in Fig.~\ref{fig:topology}(a,c) and \ref{fig:topology}(b) respectively, are relatively simple and should be fairly straightforward to construct using sandwich structures. We now show how to change the system between these two different topologies. As shown in Fig.~\ref{fig:topology}, going between the two topologies is essentially adding and removing a strip of topological fluid which forms an ``overpass.'' While it appears that one simply needs a removable part that can be glued in and cut out as desired, literally doing this is not possible for the physical structures, hence one needs a way to \emph{effectively} do it. For this, we propose to construct the system with a handle, depicted as the combined blue and red regions in Fig.~\ref{fig:overpasses}(a), as a fixed structure, and then build an array of independently controllable depletion gates on the surface of this structure which are capable of eliminating the topological fluid from the system directly beneath them. Thus, while the given physical structure is built once and for all with a fixed topology, by turning the gates on and off, one can dynamically modify the genus of the topological system by controlling the existence of the topological fluid within the ISH structure. Specifically, by turning on the gates in the red regions in Fig.~\ref{fig:overpasses}(a,b,c), the topological fluid remains only in the blue regions, realizing the topological configurations of Fig.~\ref{fig:topology}(a,b,c), respectively. Turning the gates off allows the topological fluid to fill in previously depleted regions of the fixed structure, thus allowing the system
to change between the different configurations, e.g. from Fig.~\ref{fig:overpasses}(a) to \ref{fig:overpasses}(b). With this dynamical topology changing (DTC) ability, we can now outline the routines that allow us to implement the topologically protected gates that will enable universal quantum computation. The computationally universal gate set we will specify is given by the $\pi/8$-phase, $\pi/4$-phase, Hadamard, and ${\rm C}(Z)$ gates~\cite{Boykin99}. The $\pi/4$-phase and Hadamard gates can be obtained simply by braiding quasiparticles, but the complete set of gates requires DTC operations.

As described in~\cite{Bravyi00-unpublished,Freedman06a}, one can obtain a $\pi/8$-phase gate by the following sequence of steps:
(1) change the configuration in Fig.~\ref{fig:topology}(a) to that in \ref{fig:topology}(b); (2) perform a topological charge measurement of the boundary $C_2$;
(3) perform a topological operation
called a ``Dehn twist'' along the loop $D$ twice;
and, finally, (4) return to the configuration of Fig.~\ref{fig:topology}(a). Applying a Dehn twist along a loop mathematically means cutting the surface open along the loop, rotating one of the two resulting boundary loops $360$ degrees with respect to the other, and then gluing the two boundary loops back together to reform the original surface (but with a twist). Such an operation cannot be directly carried out
in a physical system, since one cannot simply grab the topological fluid and twist it. Thus, we propose
the following method by which Dehn twists
can be simulated. Instead of performing
steps (3) and (4), i.e. two Dehn twists on loop $D$ in Fig.~\ref{fig:topology}(b) and a return to \ref{fig:topology}(a), one can equivalently change from the configuration in Fig.~\ref{fig:topology}(b) to that in \ref{fig:topology}(c). One can envision this equivalence best by comparing the three configurations on the far right of Fig.~\ref{fig:topology}. Specifically, cutting straight across the overpass in Fig.~\ref{fig:topology}(b) would produce Fig.~\ref{fig:topology}(a). But this cut crosses the loop $D$, so if two Dehn twists were applied to $D$, the same cut would instead result in Fig.~\ref{fig:topology}(c). Thus, to implement a $\pi/8$-phase gate using a DTC structure, one controls the depletion gates to change from the configuration in Fig.~\ref{fig:overpasses}(a) to that in \ref{fig:overpasses}(b), next performs a topological charge measurement of the boundary $C_2$, and then finally changes to the twisted configuration in Fig.~\ref{fig:overpasses}(c).

For the ${\rm C}(Z)$ operation, one can control the depletion gates to change from the configuration in Fig.~\ref{fig:overpasses}(a) to that in \ref{fig:overpasses}(b), next perform a topological charge measurement of $C_2$, then transport a pair of $\sigma$ quasiparticles which encode the second qubit around the loop $D$ as indicated in Fig.~\ref{fig:overpasses}(d), and finally return to the configuration of Fig.~\ref{fig:overpasses}(a). One can also implement a ${\rm C}(Z)$ gate if one has the ability to perform non-demolitional collective topological charge measurements of four quasiparticles with good precision~\cite{Bravyi06}, for example by interferometry.

The result of the topological charge measurement of $C_2$ in the above routines determines whether the described operations result in the $\pi/8$-phase and ${\rm C}(Z)$ gates, respectively, or ones that are related to these by applications of Clifford gates which may be obtained via quasiparticle braiding. This is an example of ``adaptive computation,'' where the outcomes of intermediate measurements determine -- via a
classical control computer -- the subsequent application of gate operations.

One might be concerned about the practicality of working with twisted geometries such as in Fig.~\ref{fig:topology}(c), especially since the many applications of the $\pi/8$-phase gate needed in a computation appear
to produce too much twisting for the physical constraints of the system. This is however not as severe a problem as it seems, since the encoded quantum information can be teleported into and out of DTC structures using anyonic teleportation~\cite{Bonderson08a}. Performing such anyonic teleportations for the state encoded in the twisted geometry may be rather challenging experimentally, since it involves a topological charge measurement enclosing one of the twisted boundaries.
This is the most technologically demanding step
in our protocol, but it should be realizable in ISHs.

\begin{figure}[t!]
\centering
\includegraphics[width=3.5in]{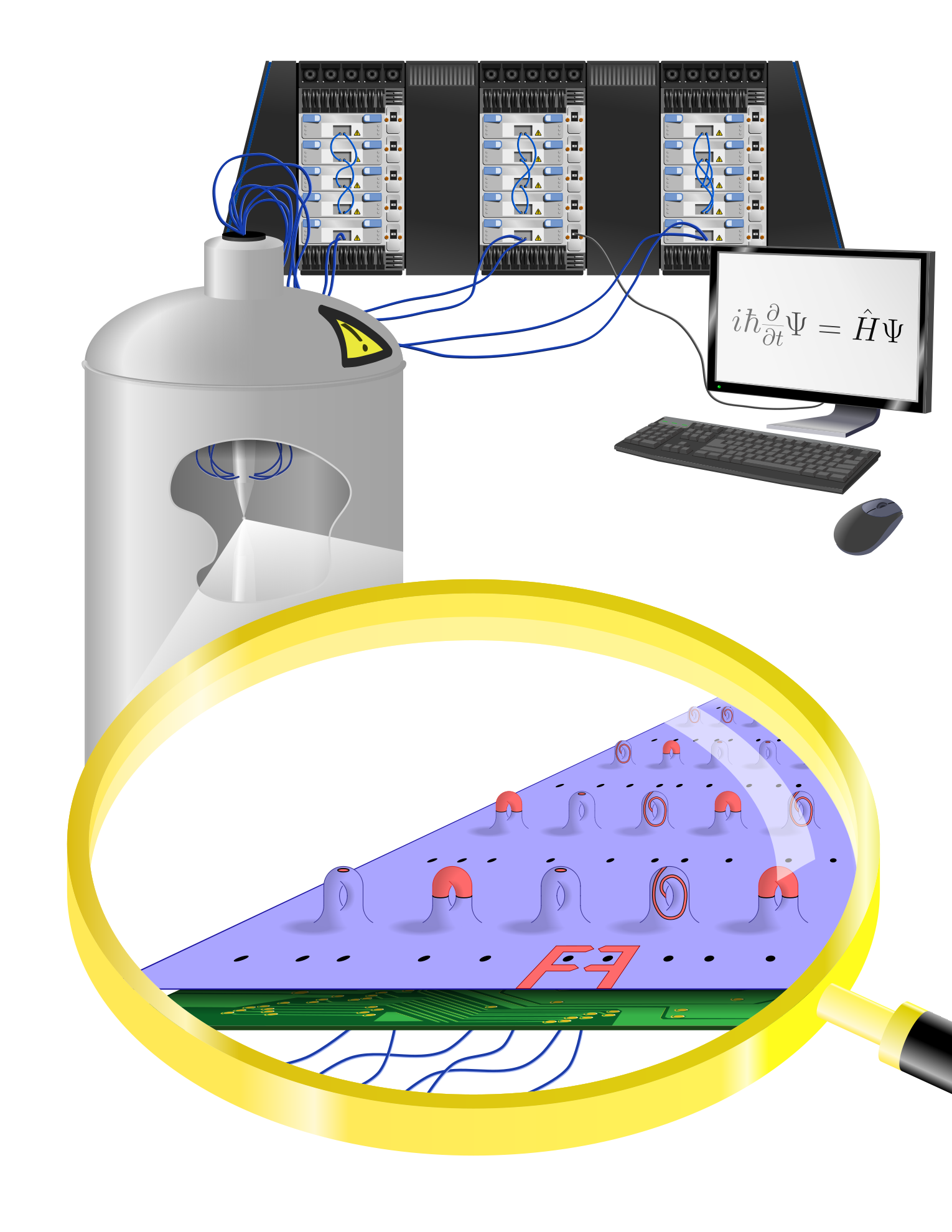}
\caption{A schematic blueprint for a universal Ising topological quantum computer:
an ISH system that is primarily planar with (non-planar) DTC structures (as in Fig.~\ref{fig:overpasses}) distributed throughout. Non-Abelian quasiparticles are represented by black dots and a topological charge measuring device (in this case an interferometer) is represented by the red ``claws'' at the front of the ISH system. The gates and devices used to perform DTC operations, quasiparticle braiding, and topological charge measurement are all controlled by a classical
computer. \newline \emph{Note: figure not drawn to scale.}}
\label{fig:quantum_computer}
\end{figure}

Thus, it is interesting to compare ISHs and the $\nu=5/2$
fractional quantum Hall state in the context
of our proposal.
The fundamental favorable characteristic of
ISH systems over 5/2 is that the energy scale of the gap is derived from the proximate superconducting gap $\Delta$ and is therefore:  (1) potentially much larger, and (2) not dependent on the realization of a high mobility heterostructure. This last requirement has, until now, dictated precise planarity for
fractional quantum Hall states. Our computationally universal proposal requires, at a fundamental level, the use of curved (non-planar) topological fluids,
which should be possible in ISHs.

We estimate the proximity-induced gap in
an ISH to be $\approx 5$~K if Nb is the superconducting element (in contrast to the $\nu=5/2$ gap of $\sim 500$~mK). This
gap, unlike in the fractional quantum Hall situation,
does not depend on ultra-high mobility since it is protected by
Anderson's theorem~\cite{Anderson59}.
Therefore, the TQC operating temperature, which
must be well below the gap, need not be
in the millikelvin range for ISH, where a dilution refrigerator
would be required. In fact, none of the microscopic design elements need to be particularly exotic.
The superconducting film should have a thickness larger than the coherence length, which is approximately
$0.1$~$\mu$m for a variety of superconductors.
In semiconductor ISHs, the semicondutor layer
should be a generic quantum well,
approximately $10-50$~nm thick, with a doping of approximately $10^{10}$ electrons/cm$^2$. The various DTC and qubit structures are on the micron scale. The gates
needed for braiding, measurement, and DTC operations are standard
metallic electrical gates and micro-magnetic
metallic gates creating local magnetic fields.

Putting all the pieces together, we draw up a schematic blueprint in Fig.~\ref{fig:quantum_computer} of an Ising anyon based topologically fault-tolerant quantum computer in which quasiparticle braiding, dynamical topology change, and topological charge measurement can all be performed, thus allowing universal quantum computation.


\end{document}